\documentclass{aa}
\usepackage{graphicx}

\usepackage[]{natbib}
\def\flux{$\cdot10^{-11}$~ph~cm$^{-2}$\,s$^{-1}$}
\def\dflux{$\cdot10^{-11}$~ph~cm$^{-2}$\,s$^{-1}$\,TeV$^{-1}$}
\def\iflux{$F_{E>1~\mathrm{TeV}}$}
\def\newcut{$3.6(+0.4-0.3)_{\rm stat}(+0.9-0.8)_{\rm sys}$~TeV}
\def\oldcut{$6.2\pm0.4_{\rm stat}(+2.9-1.5)_{\rm sys}$~TeV}
\def\dcut{$\Delta E=2.6\pm0.6_\mathrm{stat}\pm0.6_\mathrm{sys}$~TeV}

\def\bea{\begin{eqnarray*}}
\def\eea{\end{eqnarray*}}
\def\ba{\begin{eqnarray}}
\def\ea{\end{eqnarray}}
\begin{document}
\date{Received / Accepted}
\title{Variations of the TeV energy spectrum at different flux levels of Mkn~421
observed with the HEGRA system of Cherenkov telescopes}
\titlerunning{Spectral variations of Mkn~421 during flares}
\author{
F.~Aharonian\inst{1},
A.~Akhperjanian\inst{7},
M.~Beilicke\inst{4},
K.~Bernl\"ohr\inst{1},
H.~B\"orst\inst{5},
H.~Bojahr\inst{6},
O.~Bolz\inst{1},
T.~Coarasa\inst{2},
J.~Contreras\inst{2},
J.~Cortina\inst{2},
L.~Costamante\inst{1},
S.~Denninghoff\inst{2},
V.~Fonseca\inst{3},
M.~Girma\inst{1},
N.~G\"otting\inst{4},
G.~Heinzelmann\inst{4},
G.~Hermann\inst{1},
A.~Heusler\inst{1},
W.~Hofmann\inst{1},
D.~Horns\inst{1},
I.~Jung\inst{1},
R.~Kankanyan\inst{1},
M.~Kestel\inst{2},
J.~Kettler\inst{1},
A.~Kohnle\inst{1},
A.~Konopelko\inst{1},
H.~Kornmeyer\inst{2},
D.~Kranich\inst{2},
H.~Krawczynski\inst{1,9},
H.~Lampeitl\inst{1},
M.~Lopez\inst{3},
E.~Lorenz\inst{2},
F.~Lucarelli\inst{3},
O.~Mang\inst{5},
H.~Meyer\inst{6},
R.~Mirzoyan\inst{2},
M.~Milite\inst{4},
A.~Moralejo\inst{3},
E.~Ona\inst{3},
M.~Panter\inst{1},
A.~Plyasheshnikov\inst{1,8},
G.~P\"uhlhofer\inst{1},
G.~Rauterberg\inst{5},
R.~Reyes\inst{2},
W.~Rhode\inst{6},
J.~Ripken\inst{4},
G.~Rowell\inst{1},
V.~Sahakian\inst{7},
M.~Samorski\inst{5},
M.~Schilling\inst{5},
M.~Siems\inst{5},
D.~Sobzynska\inst{2,10},
W.~Stamm\inst{5},
M.~Tluczykont\inst{4},
H.J.~V\"olk\inst{1},
C.~A.~Wiedner\inst{1},
W.~Wittek\inst{2},
and R.~A.~Remillard\inst{11}}

\institute{Max-Planck-Institut f\"ur Kernphysik, Postfach 103980, D-69029 Heidelberg, Germany
\and Max-Planck-Institut f\"ur Physik, F\"ohringer Ring 6, D-80805 M\"unchen, Germany
\and Universidad Complutense, Facultad de Ciencias F\'{\i}sicas, Ciudad Universitaria, E-28040 Madrid, Spain
\and Universit\"at Hamburg, Institut f\"ur Experimentalphysik, Luruper Chaussee 149, D-22761 Hamburg, Germany
\and Universit\"at Kiel, Institut f\"ur Experimentelle und Angewandte Physik, Leibnizstra{\ss}e 15-19, D-24118 Kiel, Germany
\and Universit\"at Wuppertal, Fachbereich Physik, Gau{\ss}str.20, D-42097 Wuppertal, Germany
\and Yerevan Physics Institute, Alikhanian Br. 2, 375036 Yerevan, Armenia
\and On leave from Altai State University, Dimitrov Street 66, 656099 Barnaul, Russia
\and Now at Yale University, P.O. Box 208101, New Haven, CT 06520-8101, USA
\and Home institute: University Lodz, Poland
\and Center for Space Research, MIT, Cambridge, MA02139
}

\authorrunning{Aharonian et al.}

\date{Received / Accepted}

\offprints{\\D.Horns,
Dieter.Horns@mpi-hd.mpg.de}

\abstract{
The nearby BL Lacertae (BL Lac) object Markarian~421 (\object{Mkn~421}) at a red shift $z=0.031$ was 
observed to undergo strong TeV $\gamma$-ray outbursts
in the observational periods from December 1999 until May 2001.
 The time averaged flux level $F(E>1\mathrm{\,TeV})$ in the 1999/2000 season
was $(1.43\pm0.04)$~\flux, 
whereas in the 2000/2001 season the average integral flux increased to
$(4.19\pm0.04)$~\flux.  
  Both energy spectra are curved and well fit
by a power law with an exponential cut-off energy at $3.6(+0.4-0.3)_{\rm stat}(+0.9-0.8)_{\rm sys}$~TeV. 
The respective energy spectra averaged over each of the two time periods 
indicate a spectral hardening for the 2000/2001 spectrum. 
  The photon index changes
from
$2.39\pm0.09_{\rm stat}$ for 1999/2000 to $2.19\pm0.02_{\rm stat}$ in 2000/2001.
 The energy spectra derived for different average flux levels ranging from 0.5 to 10~\flux\,
follow a clear correlation of photon index and flux level. Generally, the energy spectra are harder for high flux levels. 
From January to April 2001 Mkn~421 showed 
rapid variability (doubling time as short as $20$~minutes), accompanied with a spectral hardening 
with increasing flux level within individual nights.
 For two successive nights (MJD~51989-51991, March 21-23,2001),
this correlation of spectral hardness and change in flux
has been observed within a few hours. 
The cut-off energy for the Mkn~421 TeV spectrum remains within the errors
constant for the different flux levels and
differs by \dcut\,
from the value determined for Mkn~501.
This indicates that the observed exponential cut-off in the energy spectrum of Mkn~421 
is not solely caused by absorption of multi-TeV photons 
by pair-production processes with photons of the extragalactic near/mid infrared background radiation.

}
\maketitle
\section{Introduction}
   The nearby BL Lac object Mkn~421 ($z=0.031$) was the first extra-galactic
source of very-high energy gamma-ray emission detected \citep{1992Natur.358..477P}.
 The source was monitored closely with the HEGRA imaging Cherenkov telescopes \citep{1999APh....10..275H} 
during the
observational periods from 1997-1998 while the source remained in a low flux state with
an average flux level of a third of the flux observed from the \object{Crab-Nebula}  and
a power law energy spectrum with a photon index of $3.09\pm0.07$ \citep{1999A&A...350..757A}.
 It is well known that 
the source exhibits time variability on a sub-hour timescale  at 
TeV-energies \citep{1996Natur.383..319G}, which 
constrains the size of the emission region ($R<10^{15}\,\mbox{cm}
(\delta/10)\,(T_{\rm var}/1\,{\rm hr})$) and indicates a relativistic
bulk motion  of the emitting plasma  with a Doppler factor
$\delta=(\Gamma-\sqrt{\Gamma^2-1}\cos\theta)^{-1}>10$ at an angle $\theta$ to the line of sight 
to avoid severe internal absorption
\citep{1998MNRAS.293..239C}.

The observed variations  of the 
source in the optical and radio energy region have been found to correlate only marginally with the TeV
flux \citep{2001AAS...199.9817N,ICRC2001...jordan}. However, the X-ray activity shows good correlation with the 
TeV flux level. During the observational periods December 1999 until May 2001, extended multi-wavelength observation campaigns 
were conducted by X-ray satellites (RXTE, Beppo\-SAX) and TeV
observatories (HEGRA, CAT and VERITAS collaborations).
  During  coordinated simultaneous observations  in January/February 2001  with RXTE, Mkn~421 
showed a soft X-ray spectrum consistent with a peak position of the 
synchrotron spectrum below $3$~keV. The X-ray flux integrated between 2-10~keV
reached a value of $1.6\cdot10^{-9}$~erg/(cm$^2$\,s) on January~31, 2001
\citep{ICRC2001...horns} where Mkn~421 showed also a high TeV flux level. 
This is the highest X-ray state for Mkn~421 reported during a pointed observation.
Interpretation of the broadband spectral energy distribution (SED) in the framework of 
synchrotron-self-Compton models is discussed in Sect.~\ref{MWL}.

 The TeV energy spectrum of Mkn~421 has been measured by several instruments 
(HEGRA, Whipple, CAT) at different flux levels of the source \citep{1999A&A...350..757A,1999ApJ...511..149K,2001ApJ...560L..45K,2001A&A...374..895P}. The observed spectral shape 
shows tentative indications for variations and a possible correlation with the overall flux level.
 However,
since the instruments have largely different energy thresholds and systematic uncertainties, a firm claim of
spectral variations of the TeV emission from Mkn~421 could so far not be made. For Mkn~501, 
observations of a spectral hardening during the 1997 outbursts in TeV emission have been claimed \citep{1999A&A...350...17D}, 
but could not be verified by other experiments \citep{2001ApJ...546..898A}.
The TeV-spectrum of Mkn~501 softened during the subsequent years following the strong outburst while
the mean flux has decreased by a factor of 10 to $1/3$ of the flux observed from the Crab-Nebula \citep{2001ApJ...546..898A}.

Mkn~421 exhibited strong activity from January to May 2001 \citep{2001IAUC.7568....3B}
with a peak diurnal flux level on March 23/24, 2001 (MJD 51991.9-51992.2) 
with a value of $(12.5\pm0.4)$~\flux above
1~TeV corresponding to $7.4$ times the flux of the Crab-Nebula. Only 6 days before reaching the peak flux, the diurnal flux 
had been as low as $(0.5\pm0.1)$~\flux, a factor of 25 lower.

We report on observations of Mkn~421 carried out with 
the HEGRA system of 5 imaging air Cherenkov telescopes. 
  Sects.~\ref{lightcurve}~\&~\ref{spectrum} contain the results on 
flux and spectral shape of Mkn~421 during the observational periods from December 1999 until 
May 2001. 
 We show clear evidence for variations of the
spectral shape of a TeV $\gamma$-ray source, getting harder with increased flux level (Sect.~\ref{sec:corr}).  
Intranight flux and spectral variations are evident in at least two nights of observations
 (Sects.~\ref{diurnal:flux}~\&~\ref{diurnal:spectrum}). The difference between cut-off energies present
in the TeV spectra of Mkn~421 and Mkn~501 is discussed in Sect.~\ref{subsection:differences}
 
\begin{table}
\caption{\label{table:obstime} 
Individual observation times, and mean zenith angles $\langle \theta \rangle$ are
listed for the different observational periods.
}

\begin{tabular}{lr|cc}
\hline
\hline
Date & Year & Obs.Time & $\langle \theta\rangle$ \\
     &        & $[hrs]$    &   $[^\circ]$ \\
\hline
December 3-11      & 1999  &  4.8 & 17.8 \\
January 31-February 15 & 2000  & 58.4 & 20.9 \\
March 3-March 12  & 2000  & 17.9 & 13.2 \\
March 24-March 29 & 2000  & 28.4 & 19.8 \\
April 23-May 5  & 2000  & 21.6 & 17.1 \\
\hline
November 24-December 6  & 2000  & 20.8 & 30.5 \\
December 28        & 2000  & 2.0  & 16.4 \\
January 17-February 5  & 2001  & 94.2 & 21.2\\
February 13-February 28 & 2001  & 63.2 & 20.5\\
March 17-March 31 & 2001  & 51.6 & 17.8\\
April 14-April 26 & 2001  &  8.1 & 20.6\\
May 9- May 23 & 2001  & 14.9 & 20.7 \\
\hline
Total         &       & 385.9& 20.3 \\
\hline
\end{tabular}
\end{table}

\section{Data selection and analyses}
\begin{figure*}
\begin{center}
\includegraphics[width=1.30\vsize,angle=90]{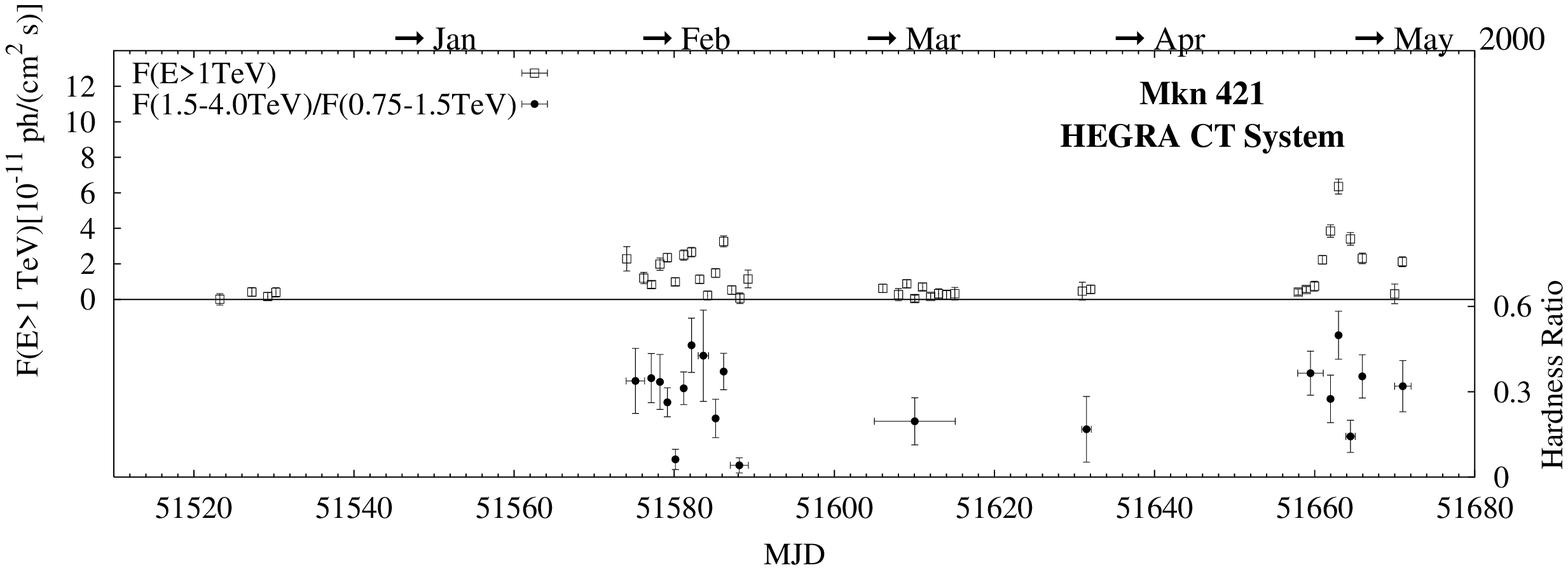}
\includegraphics[width=1.30\vsize,angle=90]{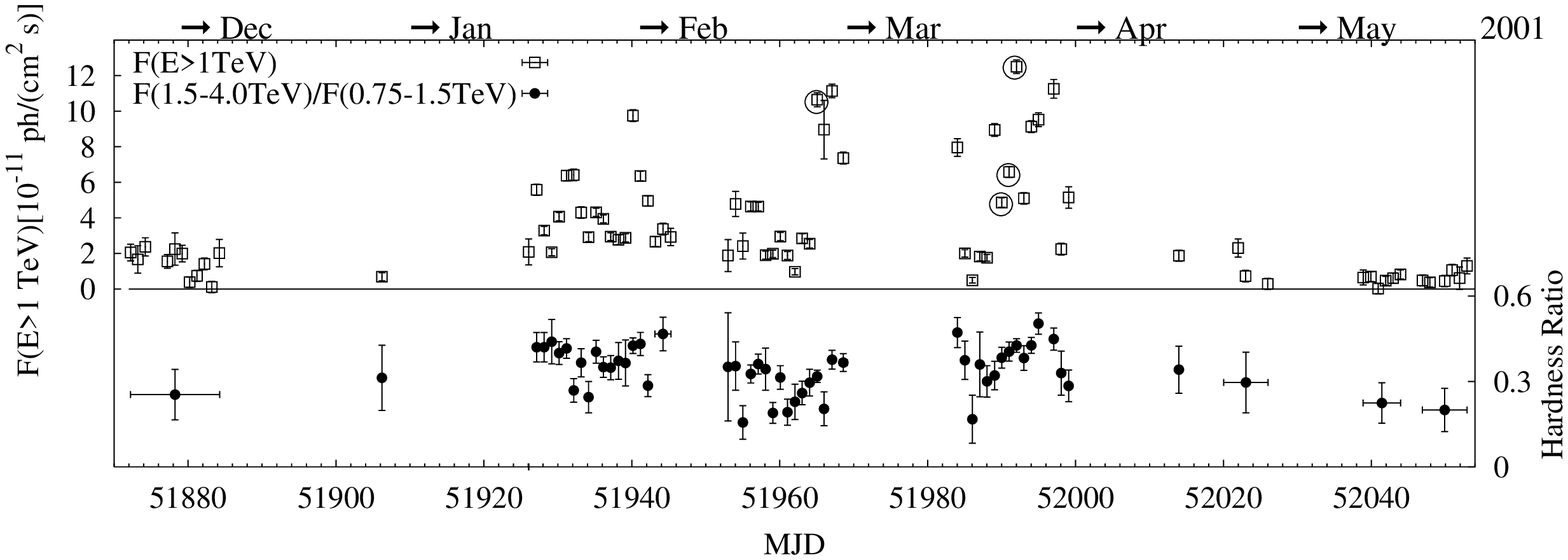}
\end{center}
\caption{\label{plot:lcurve} For the observational periods 1999/2000  and 2000/2001 
 the diurnal integral fluxes observed 
above 1~TeV are indicated. The highlighted days (circled) are subject to 
separate discussion in Sect.~\ref{diurnal:flux}\&\ref{diurnal:spectrum}. 
 The bottom part of each panel shows the hardness ratio calculated by integrating the flux in two
separate energy bands covering 0.75~-1.5~TeV and 1.5~-4.0~TeV respectively. For short diurnal exposures, consecutive
measurements have been combined.
}
\end{figure*}

\begin{figure}
 \resizebox{\hsize}{!}{\includegraphics[clip]{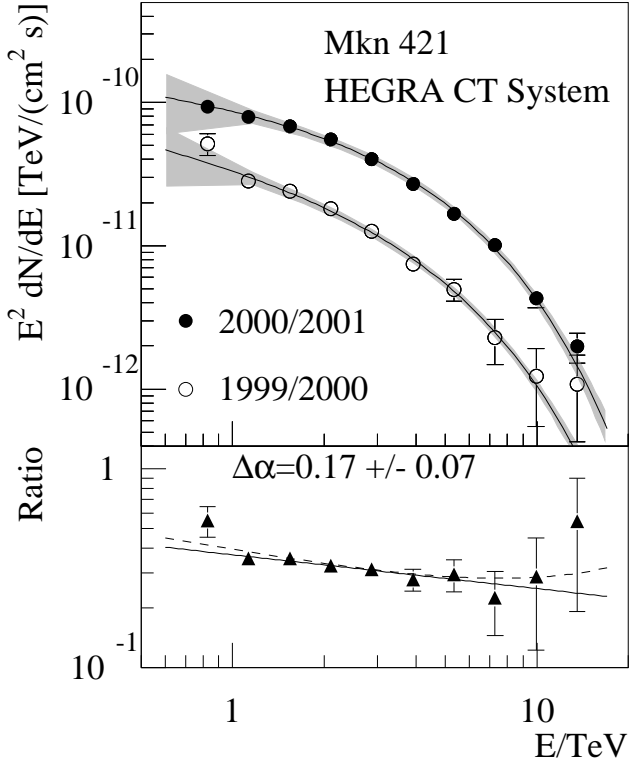}}
 \caption{\label{plot:time_averaged_spectrum} Separated into the two 
observational periods (1999/2000 and 2000/2001), 
the time averaged spectra show differences in the
slope of the spectrum being harder in 2000/2001 coinciding with an 
increase of the flux by a factor of $2.9\pm0.1$.  
  The shaded areas indicate the systematic errors on the 
spectral measurement, being most prominent in the threshold region.  Both spectra are well fit by 
a power law with an exponential
cut-off energy at $3.6$~TeV. The photon index changes by $\Delta \alpha=0.17\pm0.07$ as
indicated by a fit to the ratio of the two energy spectra which is well described by a pure power law.
The dashed line indicates a fit of a function as given in
Eqn.~\ref{eqn:ratio} with $c=(0.04\pm0.07)$~TeV$^{-1}$ indicating
that $E_0$ is consistent between the two measurements.
}
\end{figure}
\begin{figure}
 \resizebox{\hsize}{!}{\includegraphics{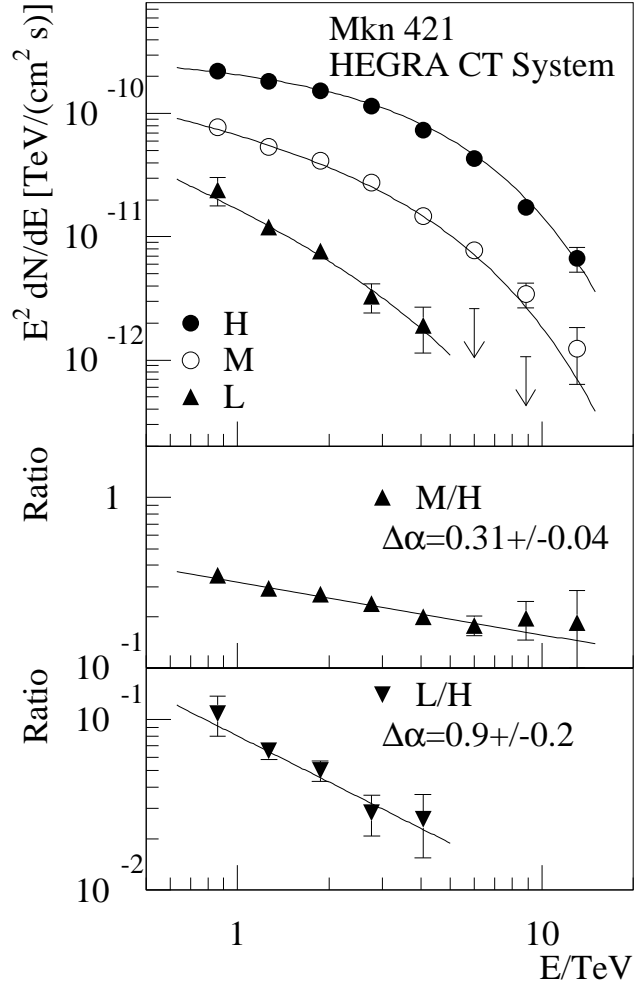}}
 \caption{\label{fig:hilo} 
 For three distinct flux intervals (\textbf{H,M,L}, see Table~\ref{table:flux}), the energy spectra have
been extracted and the flux ratios calculated. The energy spectra get considerably 
softer with decreasing average flux. The cut-off energy remains unchanged within the statistical error.}
\end{figure}

\begin{table*}
\caption{\label{table:flux}Listed are the observational time ($T_{obs}$), the integral flux above 1~TeV and the
results of $\chi^2$ minimisation on the differential photon energy spectrum using a function of 
the form $dN/dE=N_0 (E/\mbox{TeV})^{-\alpha} \exp(-E/E_0)$. The cut-off energy $E_0$ has been 
fixed to the value derived from a fit to the combined data set taken from December\,2000 until May\,2001. 
The quoted errors are statistical errors only.
The lower part of the Table summarises the result of applying fits of a power law function to the time averaged spectra and
examples of diurnal energy spectra with spectral variations during flares.
}
\small
\begin{center}
\begin{minipage}{\linewidth}
\begin{tabular}{lcccccc}
 \hline
\hline
 Period\rule{0pt}{0.4cm} & $T_{obs}$ & \iflux     &$N_0$                        & $\alpha$       & $E_0$
& $\chi^2_\mathrm{red}(d.o.f.)$\footnote{$\chi^2_{\mathrm{red}}:= \chi^2/d.o.f.$} \\[0.2cm]
        &  $[hrs]$        
& $\left[\frac{10^{-11}}{\mbox{cm}^2\,\mbox{s}}\right]$&
$\left[\frac{10^{-11}}{\mbox{cm}^2\,\mbox{s}\,\mbox{TeV}}\right]$ &    & $[$TeV$]$                  &                \\
\hline
\multicolumn{3}{l}{\rule{0pt}{0.3cm}$dN/dE=N_0 E^{-\alpha} \exp(-E/3.6\mbox{TeV})$}   & &
&Fixed $E_0$& \\
 \hline
Dec 1999-May 2000 & 131.1 & $1.43\pm0.05$ & $4.3\pm0.3$                &$2.39\pm0.09$   &3.6 & $0.8(8)$   \\ 
Nov 2000-May 2001 & 254.8 & $4.19\pm0.04$ & $11.4\pm0.3$              &$2.19\pm0.02$   &3.6 & $1.2(8)$   \\ 
$0<F_{-11}<1$\footnote{$F_{-11}=F(E>1\mathrm{TeV})/10^{-11}\mathrm{ph}\,\mathrm{cm}^{-2}\,\mathrm{s}^{-1}$} &  42.9  & $0.53\pm0.04$  & $2.2\pm0.3$      &$3.0\pm0.2$     &3.6      &   $0.3(3)$\\
$1<F_{-11}<2$ &  77.4  & $1.55\pm0.04$   & $5.2\pm0.2$       &$2.47\pm0.08$    &3.6 & $0.9(6)$\\
$2<F_{-11}<4$ &  54.1  & $2.76\pm0.05$  & $8.8\pm0.3$      &$2.36\pm0.05$    &3.6 & $0.9(6)$\\
$4<F_{-11}<8$ & 129.4  & $4.22\pm0.03$  & $12.1\pm0.2$     &$2.18\pm0.02$    &3.6 & $0.8(6)$\\
$8<F_{-11}<16$& 17.8   & $10.3\pm0.1$  & $27.3\pm0.5$      &$2.06\pm0.03$    &3.6 & $0.5(6)$\\      
MJD51964/51965 & 3.6   & $10.6\pm0.4$& $32\pm1$      &$2.36\pm0.06$    &3.6 & $0.9(6)$\\
MJD51991/51992 & 4.1   & $12.5\pm0.4$& $31\pm1$      &$2.04\pm0.05$    &3.6 & $1.2(6)$\\
\hline 
\multicolumn{3}{l}{\rule{0pt}{0.3cm}$dN/dE=N_0 E^{-\alpha}$}       &                    &                & \textit{no cut-off} &             \\
\hline
Dec 1999-May 2000         &        &               & $3.8\pm0.2$        &$3.19\pm0.04$   &                   & $2.9(8)$   \\ 
Nov 2000-May 2001         &        &               & $10.1\pm0.1$       &$3.05\pm0.01$   &                   & $53(8)$   \\ 
MJD51989/51990,preflare   & 1.9    & $3.6\pm0.3$    & $8.1\pm0.7$        & $2.9\pm0.1$   &                   &$0.5(6)$ \\
MJD51989/51990,flare      & 1.1    & $7.54\pm0.05$  & $12.0\pm1.0$       & $2.2\pm0.1$   &                   &$1.0(5)$ \\
MJD51990/51991,preflare   & 1.5    & $3.1\pm0.4$    & $8.3\pm0.8$        & $3.2\pm0.2$   &                   &$0.4(5)$ \\
MJD51990/51991,flare      & 1.0    & $10.0\pm0.7$   & $15.2\pm0.9$       & $2.5\pm0.1$    &                   &$0.7(5)$ \\
\hline
\end{tabular}
\end{minipage}
\end{center}

\end{table*}
\begin{figure}
 \resizebox{\hsize}{!}{\includegraphics[clip]{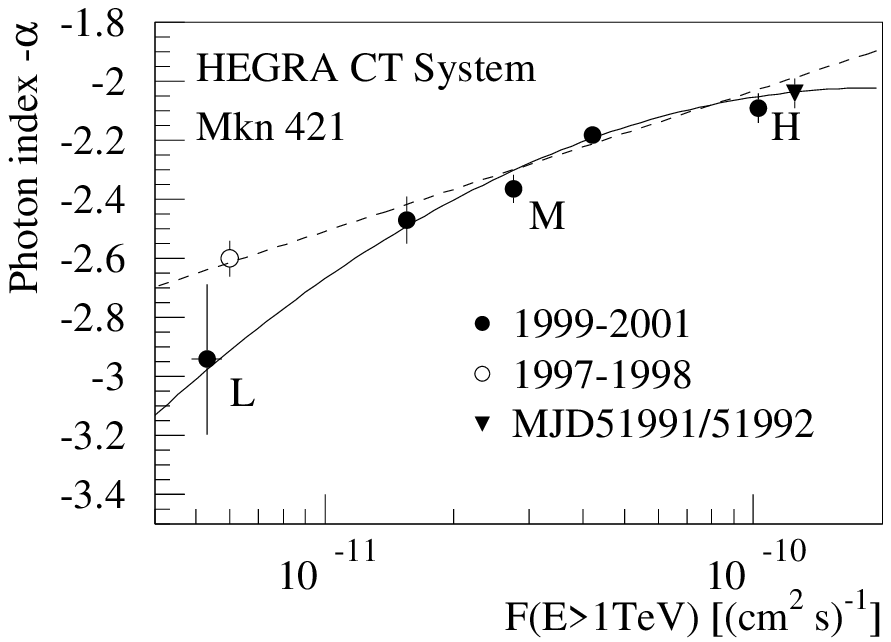}}
 \caption{\label{fig:corri}
The energy spectra get harder with increasing flux as is clearly seen when splitting
data in five separate flux intervals and fitting a power law with a fixed exponential
cut-off energy  to the individual  energy spectra. The indicated errors on the
photon index are the statistical errors.
 The solid and dashed lines indicate
 fits of  polynomial functions of the logarithm of the flux to the data (see text for details).
Also indicated are the values for the
night with the highest flux and as an open symbol 
the archival data from the 
1997/1998 observations of Mkn~421 \citep{1999A&A...350..757A}.}
\end{figure}

  The data used here were collected  from December 1999 until May 2001  and 
comprises in total 385.9~hours of observations under stable
weather conditions, good atmospheric transparency and good detector performance (see
Table~\ref{table:obstime} for a breakdown of the observation time for individual months).
Data taken up to zenith angles of $45^\circ$ are used which constitute more
than 95\,\% of the data taken in total. 

  Based upon comparing the observed rate with the expected cosmic ray event rate, individual
runs were rejected if the relative deviation from the expectation exceeds 20\,\%. In total 35\,hrs of data fail to pass this selection.\\
A large
fraction of data were taken with a 5-telescope setup. However, in January 2000
and April 2001,  only 4 telescopes were operating. The collection areas for the spectral
analysis of data taken with a 4 or 5 telescope setup are calculated 
separately to take the difference in acceptance between the two setups  into account. Additionally, the
degrading mirror reflectivity and the
conversion factors relating the registered digitised Cherenkov amplitude values in the cameras to the
number of photoelectrons detected have been determined for monthly observational
periods and are applied to the Monte-Carlo simulation to calculate the collection areas for 
each period separately. 

  All observations were carried out in the so-called \textit{wobble}-mode.
 During \textit{wobble}-mode observations the
source direction is positioned $\pm0\fdg5$ shifted in declination with respect to
the centre of the field of view of the camera. This observation mode allows for simultaneous
estimate of the background (OFF) rate induced by charged cosmic rays. 
The
analysis uses an extended OFF-region to reduce the statistical error on the
background estimate. A ring segment ($180^\circ$ opening angle), from
$0\fdg3$ to $0\fdg7$ distance from the camera centre at the opposite side
of the ON region has been chosen.    The events suitable
for spectral analysis are selected in a similar way as described in
\citet{1999A&A...350..757A}, applying only loose cuts on the image shape and
the shower direction in order to minimise systematic effects from energy
dependent cut efficiencies.  The loose directional and shape cuts accept events with 
the reconstructed direction being within an angular distance of $0\fdg22$
to the source and the image shape parameter \textit{mean scaled width} $mscw<1.2$. 
80\,\% of the registered $\gamma$-events
pass both the directional and the shape cut.  To ensure that the
images are not truncated because of the limited field of view, images with a
$distance>1\fdg7$ to the centre of the camera are rejected. At the same time
the reconstructed core position is required to be within 200~m distance to the
central telescope.  The energy reconstruction requires two images with more
than 40~photoelectrons after applying a two-stage tail-cut removing night sky
background contamination of the image. The angle subtended between the major
axes of at least two of the registered images is required to be larger than $20^\circ$.
The number of excess events after
applying all selection criteria  used in this analysis
amounts to $\approx$~40\,000, constituting the largest photon sample
collected from a single source with the HEGRA telescopes.
%
\subsection{Spectral analysis method}
 The energy
reconstruction and the reconstruction of energy spectra have been described in
detail elsewhere \citep{1999A&A...349...11A}. 
  For a large part of the data, contemporaneous
observations of the Crab-Nebula\footnote{The Crab-Nebula is frequently used 
as a standard candle for TeV astronomy} are available, allowing to compare and to verify the
expectation for detection rates and cut-efficiencies for $\gamma$-induced air
showers derived from Monte-Carlo simulations with data.

 In the possible presence of a deviation from a pure power law in the
form of an exponential cut-off, the reconstruction of the spectral shape 
beyond this cut-off is influenced by the energy resolution. Even with an event-by-event
energy reconstruction accuracy of $\Delta E/E\approx 20\,\%$ as 
achieved with the stereoscopic reconstruction technique of
air showers, a considerable 
fraction of events with overestimated energies contribute to the flux of neighbouring bins.

This becomes increasingly important for energies well in excess of the cut-off energy. 
Different methods like deconvolution using  suitable algorithms 
and forward-folding techniques are applicable and have been pursued.
 An advanced energy reconstruction technique
with an improved energy resolution of $\approx 10\,\%$ \citep{2000APh....12..207H}
has been proven to show similar results as obtained with the analysis technique used here \citep{2001A&A...366...62A}. However,
the results presented here are obtained with the conventional energy reconstruction method being less sensitive to
changes in the detector performance and therefore less susceptible to systematic errors.

The influence of the energy resolution has been absorbed in the collection area $A(E_{\rm reco},\theta)$,
$\theta$ indicating the zenith angle. The method applied here follows the same approach as described in \citet{1999A&A...349...11A}, 
using a collection area $A=A(E_{\rm reco.},\theta)$ being a function of the 
reconstructed energy $E_{\rm reco}$. The collection area therefore depends upon the spectral shape assumed. 
An iterative method
is applied to test a spectral hypothesis on the data.
 The procedure starts
with an assumed arbitrary spectral shape. For the sake of simplicity, a Crab-like power law spectrum with a 
 photon index of $2.6$ is assumed. If the reconstructed spectrum 
shows deviations from this shape, a new hypothesis can be tested by
using a different spectral shape to derive the collection area and following
the same procedure once more. The process is iterative and usually converges after
two iterations. 

 The method has been carefully checked with simulated energy spectra and also against other 
methods and has proven to be less dependent upon the exact modelling of the detector response
with Monte-Carlo methods than more sophisticated procedures as mentioned above.
\subsection{Systematic uncertainties}
\label{sect:systematics}
 The systematic errors include as most important 
contributions the uncertainties in the conversion factors used to
calculate the number of photo-electrons from the digitised pulses of the photo multiplier tubes, 
non-linearities of the electronic chain and the conservatively estimated
variations of the response of the telescopes in the threshold region. Additionally, 
the uncertainties in the spectral shape enter the calculation of the collection 
area and cause an additional systematic uncertainty especially at energies
well above the exponential cut-off energy.

 Besides the mentioned uncertainties, the overall flux determination is subject to a $15\,\%$
uncertainty in the calibration of the absolute energy scale of the 
telescopes. This uncertainty only affects the
absolute flux calibration and does not influence the reconstructed shape
of the spectrum.  The stability of the calibration of the energy scale has been
carefully tested with data taken on the Crab-Nebula constraining possible time dependent
effects seen as variations on the reconstructed flux 
to be less than $10$\% which translates into less than $6$\%
uncertainty on changes of the energy scale (assuming a power law energy spectrum with a photon index of $2.6$).
 A more in-depth discussion of the systematic errors is given in 
\citet{1999A&A...349...11A}.

\section{Results}
\begin{figure}[ht!]
\resizebox{\hsize}{!}{\includegraphics{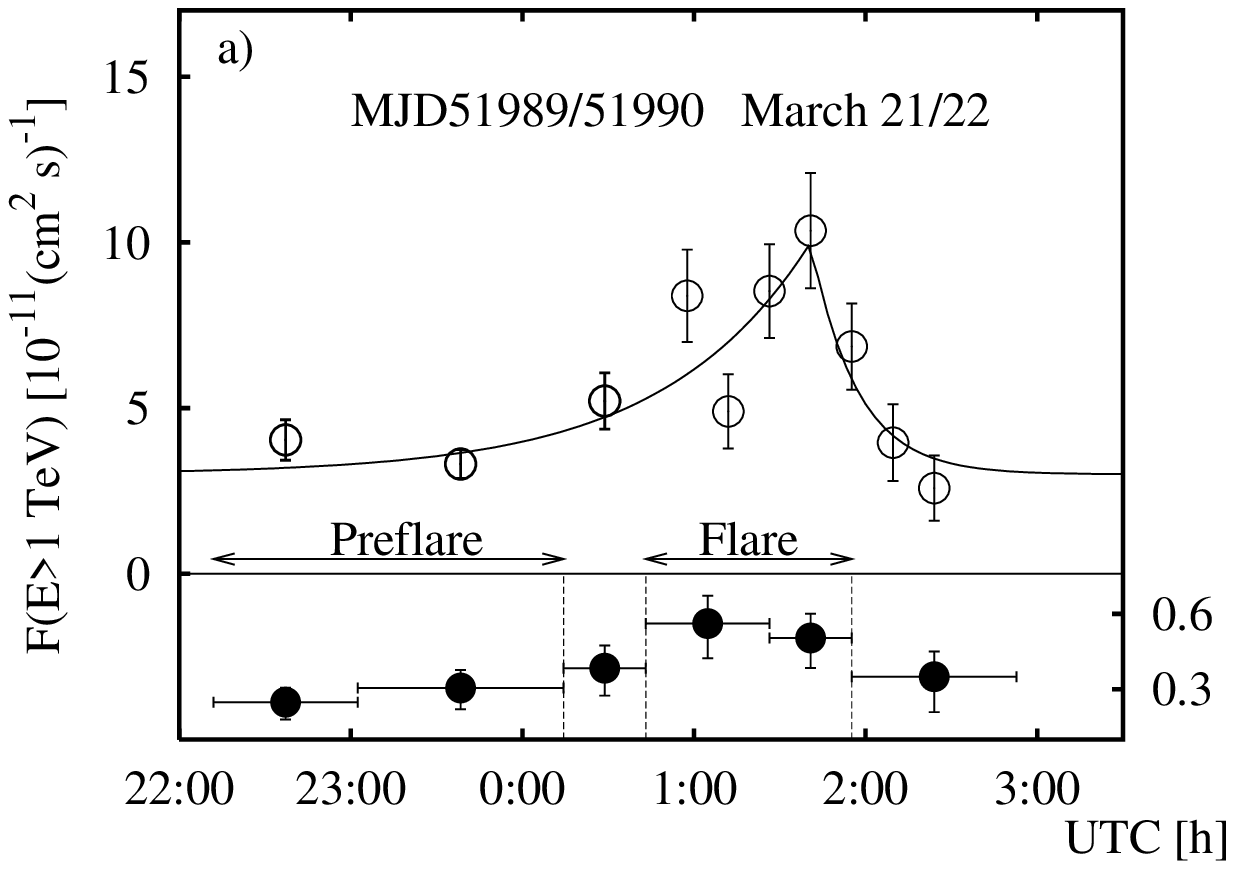}}
\resizebox{\hsize}{!}{\includegraphics{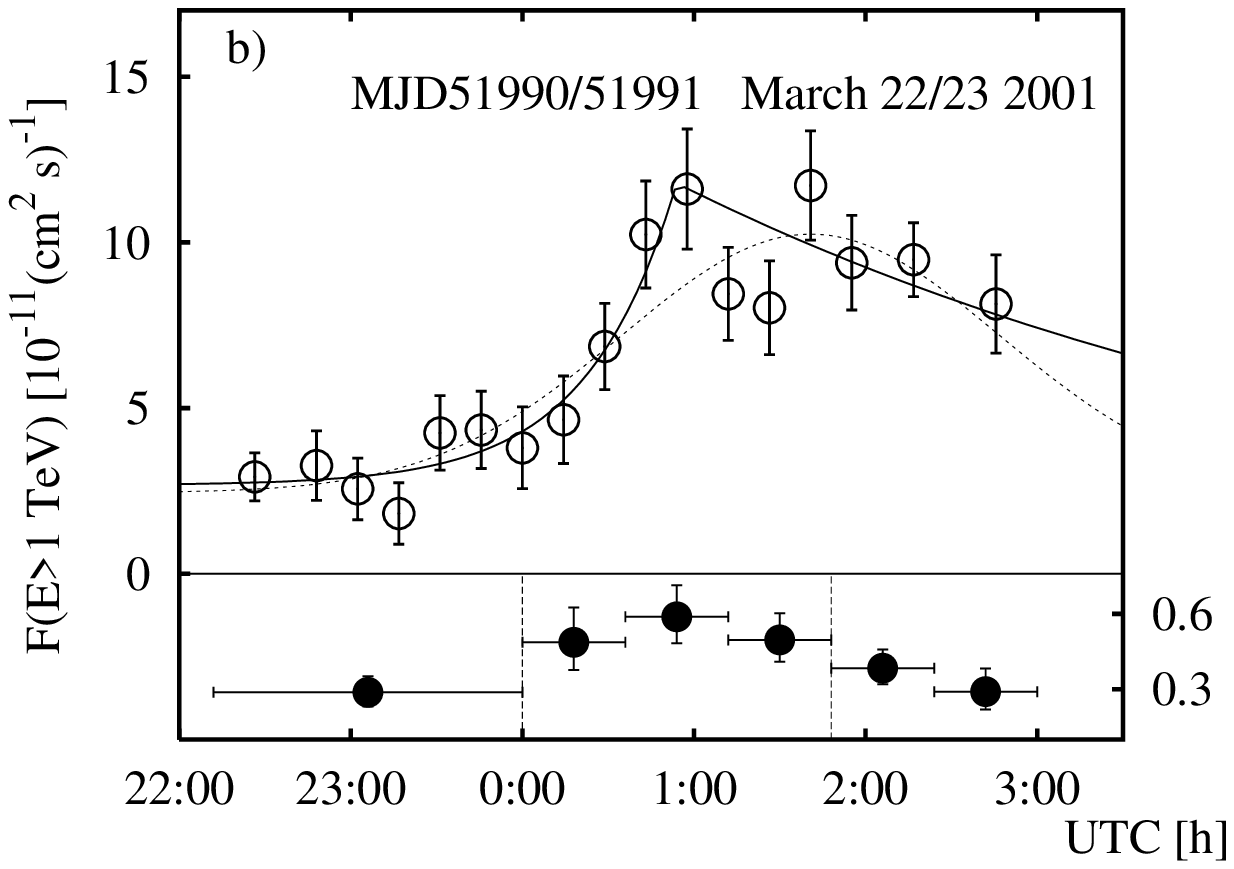}}
\caption{\label{fig:decayandrise} 
	\textbf{a)} The light curve includes the rise and decay
		of a flare observed during the night March 21/22, 2002. The solid line
		indicates a fit of a function $F(t)$ as given in Eqn.~\ref{eqn:lcurve}.
		The hardness ratio shows a hardening of the
		spectrum correlated with the increased flux.\,\, \textbf{b)}
	The lower figure shows the fastest rise time observed in the data. A fit of an exponential 
		profile (Eqn.~\ref{eqn:lcurve})  to 
		the light curve indicates a doubling time of $21\pm2$~minutes and a decay time of $\approx4.5$~hrs. However, 
	the sensitivity to a specific profile of the flare is limited by the accuracy of the flux measurement.
		A Gaussian function (dashed curve) is also an acceptable description of the data.
		Again, as in Fig.~\ref{fig:decayandrise}\textbf{a}, the 
		energy spectrum shows a hardening and softening correlated with the increase and decrease of the flux.
}
\end{figure}

\subsection{Light curve}
\label{lightcurve}
The observed integral flux above an energy of 1~TeV for individual nights is shown in
Fig.~\ref{plot:lcurve}. 
In the year 2000, the highest diurnal flux level 
reached $(6.4\pm0.3)$~\flux (MJD~51662/51663). Note the
smooth increase and decrease of the flux level in the course of a few days observed
from MJD~51659 until MJD~51665 (May 2000). The flux averaged from December 1999 to May 2000
is $(1.43\pm0.04_{\rm stat}\pm0.2_{\rm sys})$~\flux. 
In the observational period from November 2000 until May 2001 the average flux 
and the amplitude of variations increased. The maximum diurnally averaged flux 
was observed in MJD 51991/51992 (March 23/24 2002) with $(12.5\pm0.4)$~\flux (see also Sect.~\ref{diurnal:spectrum}). The time averaged flux 
for the 2000/2001 measurements is $(4.19\pm0.04_{\rm stat}\pm0.4_{\rm sys})$~\flux, which corresponds
to 2.4 times the flux of the Crab-Nebula (the integral flux above 1~TeV
		observed from the Crab-Nebula is $(1.76\pm0.06_{\rm stat}\pm0.51_{\rm sys})$~\flux \citep{2000ApJ...539..317A}).

		\subsection{Time averaged energy spectra}
		\label{spectrum}
		The observational periods 1999/2000 and 2000/2001 are treated separately for two reasons:
		The average flux level and the amplitude of 
		variability is largely different for the two seasons  
		(see Fig.~\ref{plot:lcurve}). Secondly, the two measurements are separated by one year and
		allow to investigate secular changes by comparing the energy spectra for the two observational seasons 

		A fit of a power law function $dN/dE=N_0 (E/\mbox{TeV})^{-\alpha}$ to the energy spectrum
		derived from the observations in the years 1999/2000 results in 
		$N_0=(3.8\pm0.2_{\rm stat}\pm0.4_{\rm sys})$~\dflux, $\alpha=3.19\pm0.04_{\rm stat}\pm0.04_{\rm sys}$
		with  $\chi^2_\mathrm{red}(d.o.f.)=2.9(8)$. 
		This makes a power law fit seem unlikely given a 
		probability of 
		$5\cdot10^{-3}$ for a larger $\chi^2$-value. The soft spectral shape is
consistent with previous measurements \citep{1999A&A...350..757A}.
A fit of a power law with an  exponential cut-off 
\ba
	\label{eqn:fit}
\frac{dN}{dE} &=& N_0 \cdot \left(\frac{E}{\mathrm{TeV}}\right)^{-\alpha}\cdot \exp\left(-\frac{E}{E_0}\right)
	\ea
	results in a fit to the data with $N_0=4.3\pm0.3_\mathrm{stat}\pm0.5_\mathrm{sys}$~\dflux,
	$\alpha=2.5\pm0.1_\mathrm{stat}\pm0.04_\mathrm{sys}$, $E_0=3.8\,{+5\choose -1}_\mathrm{stat}\,{+0.9 \choose -0.8}_\mathrm{sys}$~TeV
	and $\chi^2_\mathrm{red}(d.o.f.)=0.9(7)$. The large statistical error on the position of the cut-off is 
	a result of the strong correlation of the two parameters ($\alpha,E_0$) used to characterise the spectral shape. 

	The energy spectrum derived from observations in the period from November 2000 until May 2001 with 
	a substantially larger number of photons and therefore reduced statistical fluctuation is not consistent
	with a pure power law function ($\chi^2=424$ with 8 degrees of freedom). A fit of a power law function with an
	exponential cut-off results in
	a smaller $\chi^2$-value of $8.8$ with 7 degrees of freedom and 
	$N_0=11.4\pm0.3_\mathrm{stat}\pm0.4_\mathrm{sys}$, 
	$\alpha=2.19\pm0.02_\mathrm{stat}\pm0.04_\mathrm{sys}$, 
	$E_0=3.6\,{+0.4\choose-0.3}_\mathrm{stat}\,{+0.9\choose -0.8}_\mathrm{sys}$~TeV.
	Table~\ref{table:flux} shows the individual results on the power law fit and 
	fitting a power law with an exponential cut-off to different time
	periods and flux levels. 

	The energy spectra derived individually from the two observational 
	periods are presented in Fig.~\ref{plot:time_averaged_spectrum}. 
	The differential flux is multiplied by $E^2$ to emphasise the subtle differences between the two spectra\footnote{This
		corresponds to $\nu\,F(\nu)$:
			1~TeV/(cm$^2$ s)$=$1.6~erg/(cm$^2$ s).}. 
			Provided that
			the exponential cut-off is at the same energy for both years ($E_0=3.6$~TeV), the ratio of the two spectra should
			follow a power law. The lower panel of Fig.~\ref{plot:time_averaged_spectrum} shows the ratio of the two energy
			spectra.  Systematic effects that apply to both data sets in a similar manner  cancel out for the ratio of the 
			two spectra.
			As indicated with a solid line in the lower panel of Fig.~\ref{plot:time_averaged_spectrum}, 
			a power law fit is a good description of the ratio. The photon 
			indices differ by $\Delta\alpha=0.17\pm0.07$ indicating 
			a trend for
			the spectrum to become harder at a higher
			flux level (see also Sect.~\ref{sec:corr}) .  The dashed curve
			is the result of a fit with an exponential 
			as expected for different cut-off energies:
			\ba
			\nonumber  f_i     &:=& \frac{dN_i}{dE} \\
			\nonumber          &=& N_i\cdot\left(\frac{E}{\mathrm{TeV}}\right)^{-\alpha_i}\cdot
			\exp(-E/E_i)\\
			\label{eqn:ratio}
			f_1/f_2 &\propto& E^{-\Delta\alpha}\cdot\exp(-c\cdot E) \\
			\nonumber  \Delta\alpha &=& \alpha_1-\alpha_2\\
			\nonumber  c     &=& E_1^{-2}-E_2^{-2}
			\ea
	The result of a $\chi^2$-fit to $f_1/f_2$ (see Fig.~\ref{fig:hilo}) results in a difference
$c=(0.04\pm0.07)$~TeV$^{-1}$ ($\Delta E=(0.5\pm0.8)$~TeV) 
	which is consistent with the two cut-off energies being identical. 

	\subsection{Correlation of flux and photon index}
	\label{sec:corr}
	To investigate possible correlations of the flux level and the spectral shape, the data have been split 
	in five flux intervals as indicated in Table~\ref{table:flux}. 
	The energy spectra for three of the flux bins (labelled \textbf{L}ow, \textbf{M}edium, and 
			\textbf{H}igh)  are shown in Fig.~\ref{fig:hilo}. The different flux levels 
	show different spectral slopes. The ratio of the energy spectra is again 
	well-described by a power law indicating that the cut-off energy for the different flux levels remains constant 
	within the statistical error,  
	whereas the photon index changes by $\Delta\alpha=0.31\pm0.04$ for the ratio of the
	\textbf{H} and \textbf{M}. For an average flux of $F_{-11}<1$ (\textbf{L}ow), the spectrum softens such
	that the spectral index changes by $\Delta \alpha=0.9\pm0.2$  with respect  to \textbf{H}.  

	The general trend of hardening of the energy spectrum with increased flux
	is shown in Fig.~\ref{fig:corri} where the photon index $\alpha$ is shown 
	as a function of integral flux above 1~TeV.  To parameterise the 
	dependence of photon index and flux, a function of the
	form 
	\bea
	\alpha(\mbox{\iflux})& = & \sum_{i=0}^2a_i\,\log^{i}_{10}(\mbox{\iflux}) \\
	a_0 &=& -2.7\pm0.1 \\
	a_1 &=& 1.1\pm0.3 \\
	a_2 &=& -0.4\pm0.2\\
	\eea
	with $\chi^2_\mathrm{red}(d.o.f.)=1.1(2)$  represents the data points quite well. 
	A simpler functional dependence of the form
	\bea
	\alpha(\mbox{\iflux}) &=& b_0+b_1\cdot \log_{10}(\mbox{\iflux})\\
	b_0 &=& 2.7\pm0.1 \\
	b_1 &=& 0.47\pm0.02 
	\eea
	is compatible with the data  with a $\chi^2_\mathrm{red}(d.o.f.)=2.3(3)$ 
	(indicated by the dashed curve in Fig.~\ref{fig:corri}).  
	A $\chi^2$-fit of a constant value is excluded by a very large $\chi^2_\mathrm{red}(d.o.f.)=14(4)$.

	It is interesting to note that the energy spectrum of Mkn~421 observed during 1997 and
	1998 \citep{1999A&A...350..757A} fits nicely in this picture when applying the
	same fitting method to the archival data keeping the cut-off energy fixed at $3.6$~TeV (see
			Fig.~\ref{fig:corri}). This data-point has been excluded from the fitting of the functional dependence
	of the photon index.

	\subsection{Diurnal flux variations}
	\label{diurnal:flux}

\begin{figure}
\resizebox{\hsize}{!}{\includegraphics{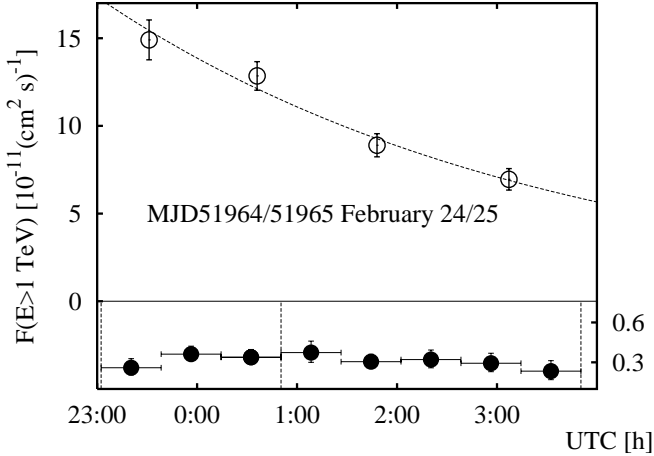}}
\caption{\label{fig:diurnal1}
The lightcurve measured over the course of 3.6\,hrs shows the decaying part of a flare that probably started before
observations began. This light curve 
 is well described by a simple exponential function $F(t)=f_0\,\exp(-t/\tau_{decay})$ with
 a decaying time $\tau_{decay}=4.5(+0.5-0.4)$~hr.
The hardness ratio (lower part of the figure) for this night remains constant
within the errors and the low value of the hardness ratio indicates a soft spectrum despite the
high flux level.
} 
\end{figure}

 Intra-night variations of the flux are seen during a large fraction of the observational nights.
In the past, doubling times as short as 15 minutes have been observed \citep{1996Natur.383..319G}. 
 To quantify the rise and decay times of flares, a simple profile 
consisting of two exponentials with a constant background is fit to the
light curves:
\ba
\label{eqn:lcurve}
 F(t) &=& f_0 + f_1\cdot\left\{ 
  \begin{array}{ll}
  \exp(\tau_\mathrm{rise}\cdot (t-t_0)) & t\le t_0\\
  \exp(-\tau_\mathrm{decay}\cdot (t-t_0))&  t>t_0\\
 \end{array}
 \right.
\ea
 An example of a night with a fast decay time (March 21/22 2001, MJD\,51989.93-51990.14) is shown in Fig.~\ref{fig:decayandrise}a. 
The exponential decay time as given by the result of a $\chi^2$-fit  
of a function $F(t)$ 
(Eqn.~\ref{eqn:lcurve}) to the light-curve results in
$\tau_\mathrm{decay}=15(+9-3)$~min, whereas
the rise time turns out to be $\tau_\mathrm{rise}=46(+60-15)$~min. This
result is indicating an asymmetry of the light curve during a flare. 
However, the presence of a possible 
sub-structure in the light curve  (at $\approx$1:00 UTC) which is not
resolved could be responsible for an overestimated rise time. 
The night with 
the fastest rise time (March 22/23 2001, MJD\,51990.925-51991.12)  
of the data set shows a smooth increase of the flux by a factor of 5 within 3 hours (see Fig.~\ref{fig:decayandrise}b). 
Fitting again a function $F(t)$ to the light-curve 
 results in $\tau_{rise}=0.52\pm0.03$~hr. This translates into a doubling time
of $21\pm2$~minutes. The decaying part of the light curve shows indications for a decay time of $4.5$~hr. However, 
a fit of a Gaussian function with a constant background (dashed 
line in Fig~\ref{fig:decayandrise}b) describes
the data equally well and would increase the estimated doubling time to $\approx 30$~min. 

 Another example of a diurnal light curve is shown in Fig.~\ref{fig:diurnal1} (February 24/25 2001). 
The light curve of this night is well described by a simple exponential decay with $\tau_{decay}=4.46(+0.54-0.39)$~hrs  similar to the decay time seen in the night  March 23/24 2001 (see 
Fig.~\ref{fig:decayandrise}b).
It is conceivable that the observed light curve in this
particular  night covers the  
end of a flare for which the first part was not visible for HEGRA. Another peculiarity of this night is the small 
and constant hardness ratio indicating a soft spectrum at a comparably high flux level (see also next Sect.).

\begin{figure*}
\resizebox{0.5\hsize}{!}{\includegraphics[clip]{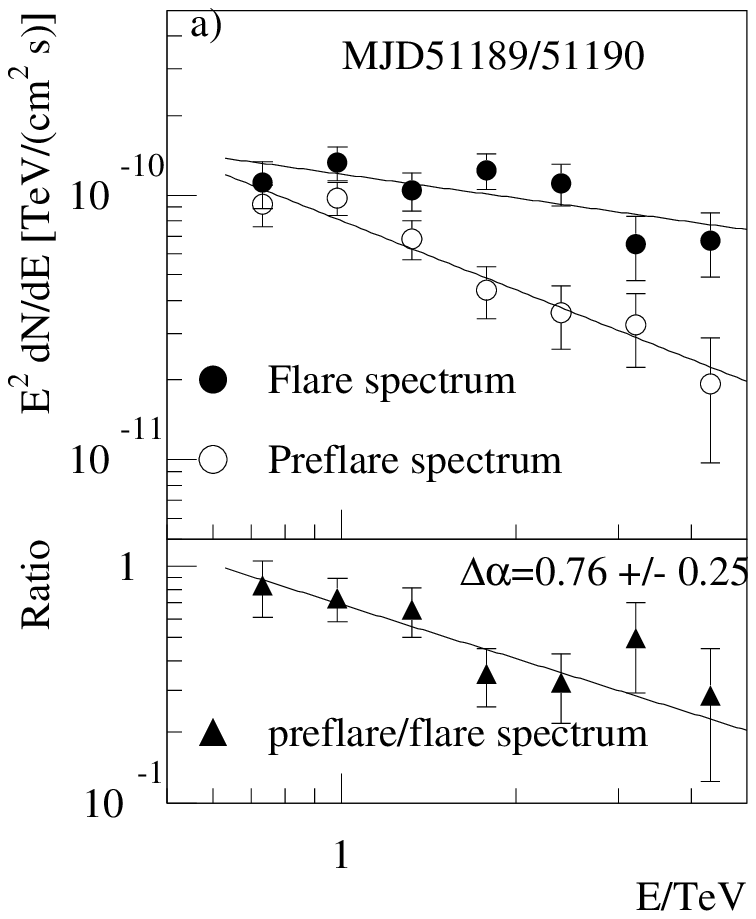}}
\resizebox{0.5\hsize}{!}{\includegraphics[clip]{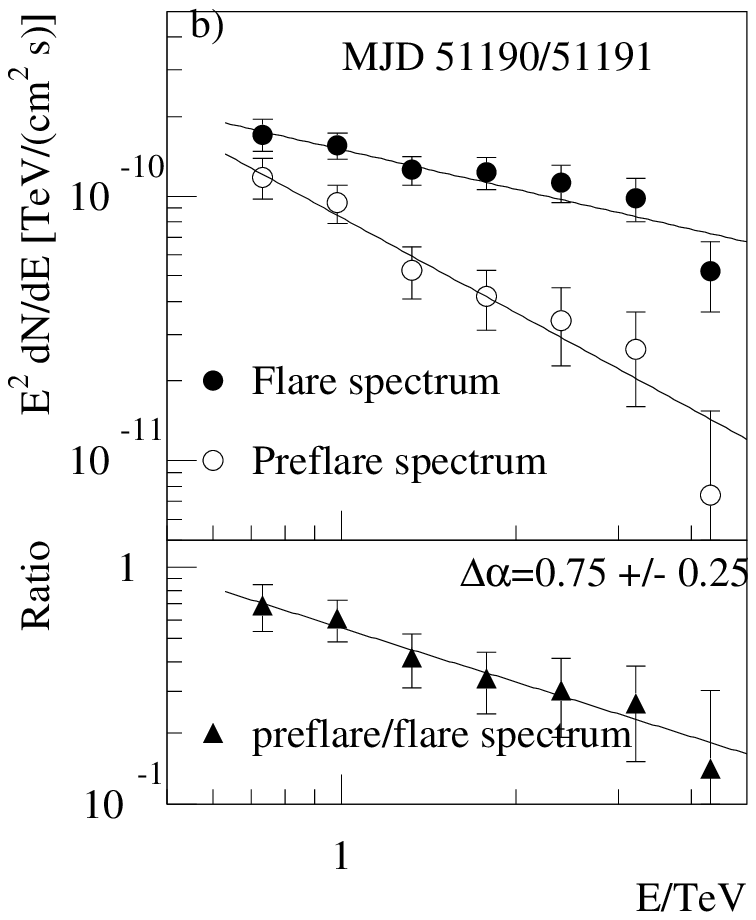}}
\caption{
\textbf{a)-b)} 
The two nights from March 21/22 (MJD51989.9-51990.2) and March 22/23 (MJD51990.9-51991.12) 
show a spectral hardening while the flux increases. For both nights, preflare and flare spectra
have been extracted and compared directly. In both cases, the photon index changes by $\Delta\alpha\approx 0.75$
while the flux increases by a factor of 3-4.  
\label{fig:1989_sp}}
\end{figure*}

\begin{figure*}
\resizebox{0.5\hsize}{!}{\includegraphics[clip]{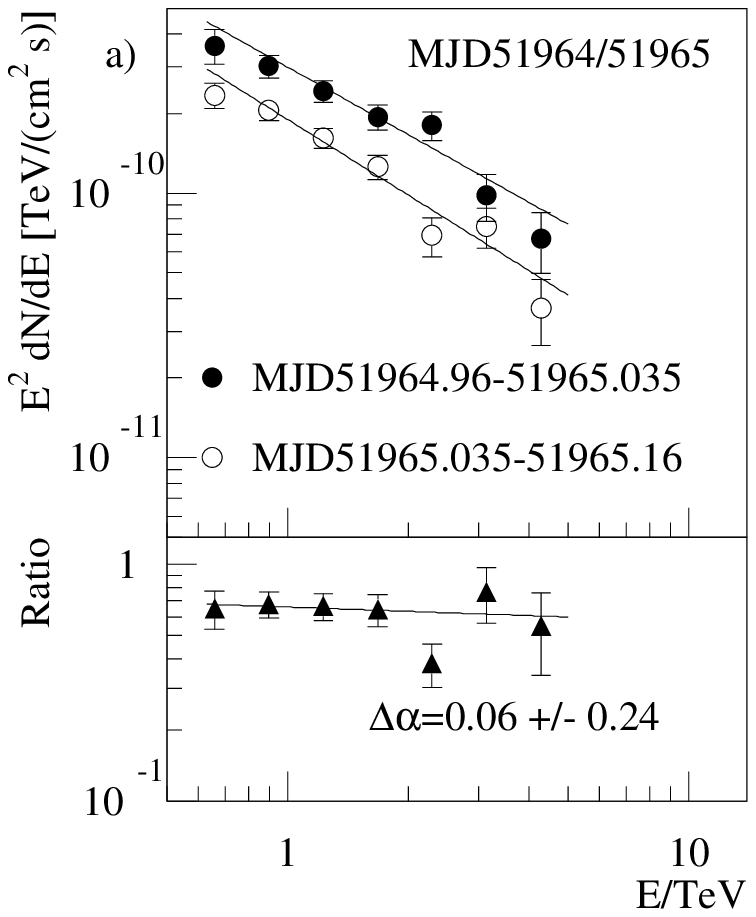}}
\resizebox{0.5\hsize}{!}{\includegraphics[clip]{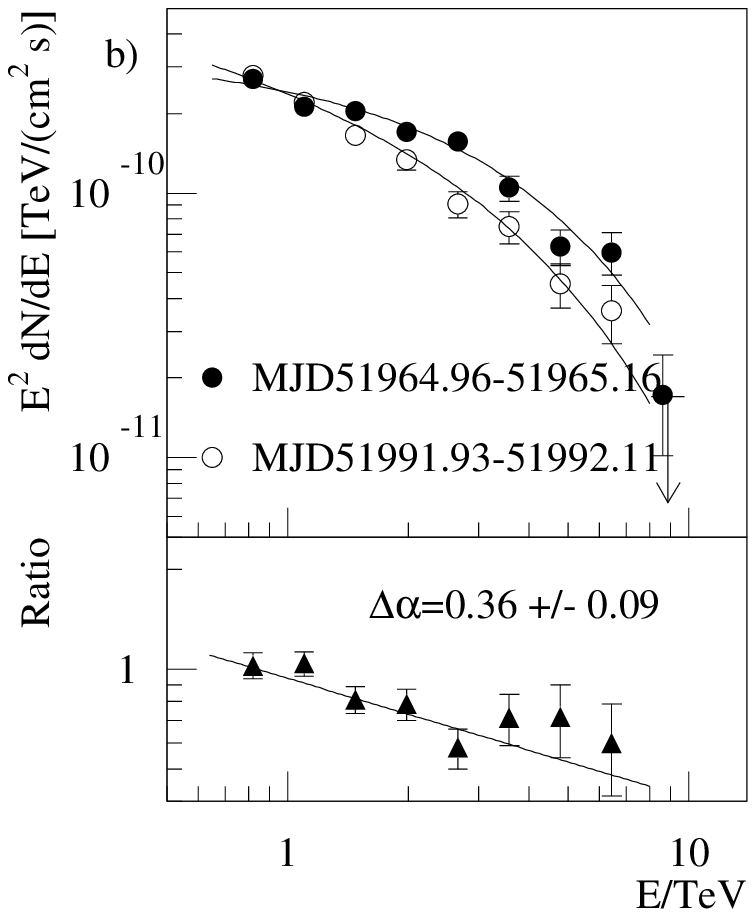}}
\caption{\label{fig:1964_sp} \textbf{a)}
For the night February 24/25, two energy spectra at different flux 
level have been extracted. No indication for a spectral evolution is visible. Both energy spectra are well
described by a pure power law with a soft photon index of $2.85\pm0.09$ and $2.95\pm0.09$ respectively. 
\textbf{b)}
 To compare two energy spectra at similar flux level but with different spectral shape, 
the energy spectra were extracted  for the night with the highest diurnal flux (March 23/24, MJD~51991/51992) and
compared with the diurnal spectrum from February 24/25. 
}
\end{figure*}

\subsection{Diurnal spectral variations}
\label{diurnal:spectrum}
Even with the large number of photons detected,
 the sensitivity for intra-night spectral changes  is limited. In a few cases, the observational time
and the number of detected $\gamma$-rays is sufficient to perform spectroscopy on time scales of hours. A particular set of
flares already mentioned in the previous Sect. 
occurred during the night MJD~51989.9-51990.2 (March 21/22 2001) (see Fig.~\ref{fig:decayandrise}a)  and the subsequent night
MJD~51990.9-51991.2 (March 22/23 2001) (see Fig.~\ref{fig:decayandrise}b). 
The light curves show in both cases a rising flux with a doubling time of less than 1\,hr.
 At the end of the flare, the flux smoothly decreased. 
During the rising time of the flux, 
the hardness ratios exhibit in both cases a spectral hardening, well correlated
with the increase of flux and a spectral softening at the end of the flare. For the night March 21/22
 the hardness ratio as well as the flux level return after the flare to the initial state it had prior to the flare. 
For the night March 22/23, the spectrum softens to the level
it had before the flare, whereas the flux remains at a high level at 
a factor of 3 higher than before the flare. 

To gain further information on the spectral shape
for the individual nights with indications for spectral variations, 
energy spectra were extracted for two time intervals  (indicated by the vertical lines in 
Fig~\ref{fig:decayandrise}a-b) from each night. 
The time intervals were chosen to cover the preflare and the flare state.
For both flares, the preflare energy spectrum is well described by a power law with a photon index of 
$\approx3.0$, and a hardening by $\Delta \alpha=0.75\pm0.25$ during the flare (see Fig.~\ref{fig:1989_sp}a-b).

However, light curves with flux variations have been observed that are not accompanied by spectral variations. An example of a night with a varying flux and a constant spectral shape is given in Fig.~\ref{fig:1964_sp}a, where
the energy spectra for two time intervals from the night MJD~51964/51965 have been extracted (marked by vertical lines in
Fig.~\ref{fig:diurnal1}). The spectrum is despite the 
large flux ($F(E>1~\mathrm{TeV})=(10.6\pm0.4)$~\flux)
comparably soft with a photon index of $2.85\pm0.09$ for the first half of the night and $2.95\pm0.09$ for
the second half of the night.

It 
is also interesting to note after comparing individual nights,  that the spectral shape  does not always relate directly to
the absolute flux-level. As an example, 
 Fig.~\ref{fig:1964_sp}b displays the spectrum of the night with the highest diurnal flux 
(MJD~51991/51992) with a level only 20\,\% higher than 
the night February 24/25 discussed earlier. During this night, no strong indication
for flux variability or changes in the hardness ratio are evident. The energy spectrum is 
hard with a photon index of $\alpha=2.04\pm0.05$ (Fig.~\ref{fig:1964_sp}b). The ratio of the two spectra indicate
a difference of $\Delta \alpha=0.36\pm0.09$. 


\section{Mkn~501 and Mkn~421 cut-off energies}
\begin{figure}
\resizebox{\hsize}{!}{\includegraphics{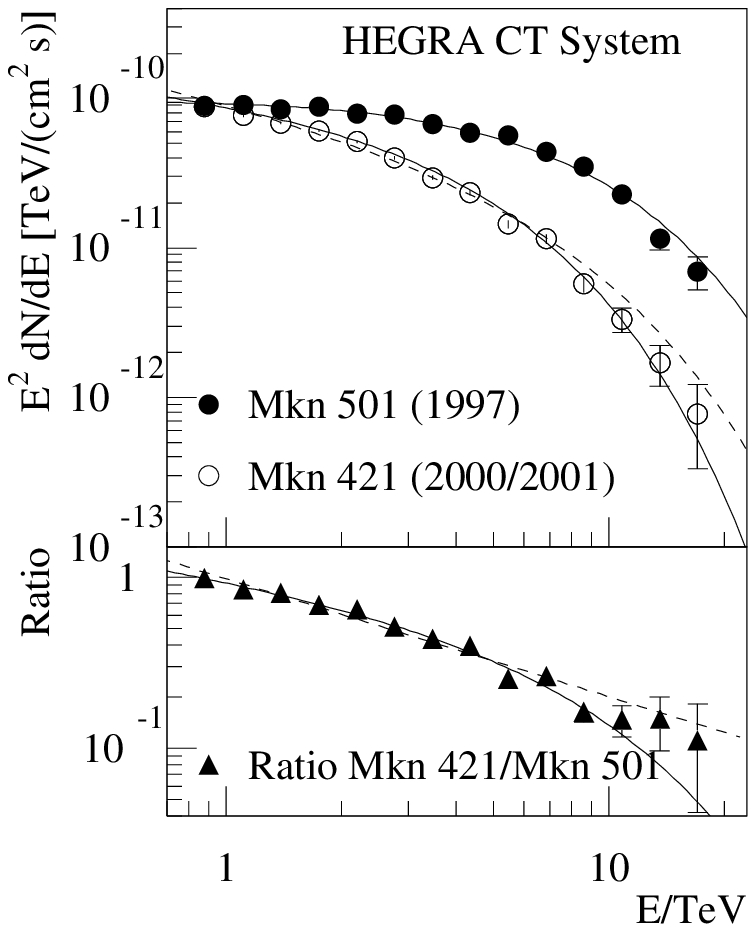}}
 \caption{\label{fig:mkn501_comp}
    Comparing the energy spectra of Mkn~501 \citep{1999A&A...349...11A} with the
measurements of Mkn~421  (2000/2001):
 The observed differential spectra are well described
by a power law with exponential cut-off (see Eqn.~\ref{eqn:fit}).
Both, the cut-off energy and the photon index differ for the 
two energy spectra ($\alpha_{Mkn501}=1.92\pm0.03_{\mathrm{stat}}\pm0.2_{\mathrm{sys}}$, $\alpha_{Mkn421}=2.19\pm0.02_{\mathrm{stat}}\pm0.04_{\mathrm{sys}}$).
A fit  of a fixed cut-off energy at 6.2~TeV as measured for Mkn~501 while letting the photon index vary
to the Mkn~421 data results in the  dashed curve in the upper panel with
$\chi^2_{\mathrm{red}}=5.7(12)$, excluding the larger cut-off energy provided that systematic effects are negligible.
To further study the difference of the energy spectra and to reduce the impact of
possible systematic effects, the ratio of the two energy spectra are calculated. 
The flux ratio is not compatible with a pure power law model 
(dashed curve, $\chi^2_{\mathrm{red}}=3.6(12)$) indicating a difference in the shape of the two energy spectra.  A function of the
form given in Eqn~\ref{eqn:ratio} gives an acceptable fit ($\chi^2_\mathrm{red}(d.o.f.)=1.5(12)$) 
with $\Delta\alpha=0.35\pm0.07$ and $c=(0.12\pm0.02)$~TeV$^{-1}$. 
}
\end{figure}

\label{subsection:differences}

 In the presence of extragalactic background light (EBL) at wavelengths of $\lambda=0.5\ldots30\,\mu$m, 
TeV-energy photons are subject to absorption by pair-production processes with the low energy
background photons \citep{1966PRL...16...252,1992ApJ...390..L49}.
As a result, the observed spectrum of extragalactic TeV sources is modified. A very likely feature 
in the observed spectrum is an exponential cut-off. 
 However, the presence of
a cut-off in the observed spectra could also be caused by  processes related to the production of TeV photons 
(e.g. decrease of the Compton-scattering cross-section for large centre-of-momentum energies, 
absorption inherent to the source). In the simplest picture of a source spectrum that follows a power law, 
 the position of the cut-off should be identical for all sources at the same red-shift provided that the 
EBL is isotropical. 

 The two well-established extragalactic sources of TeV-radiation (Mkn~421 \& Mkn~501) happen to almost coincide in red shift, which
makes an investigation of differences in the position of the cut-off energies meaningful.

   In Fig.~\ref{fig:mkn501_comp} the energy spectrum of Mkn~501 as measured during 1997 with a cut-off energy of \oldcut\, and
the energy spectrum derived from observations of Mkn~421 in 2000/2001 are compared. 
The upper panel contains the
differential energy spectra multiplied by $E^2$ together with the respective fit functions indicated as solid lines. The dashed curve is the result
of a fit of a power law with a fixed exponential cut-off at $E_0=6.2$~TeV to the Mkn~421 data points. Ignoring possible
systematic errors, the  resulting $\chi^2_\mathrm{red}(d.o.f.)=5.7(12)$ 
excludes $E_0=6.2$~TeV as a cut-off energy for the Mkn~421 spectrum.
  The lower panel of Fig.~\ref{fig:mkn501_comp} shows the ratio of the two energy spectra. A pure power law
fit to the data is very unlikely because of a 
large $\chi^2_\mathrm{red}(d.o.f.)=3.6(12)$ whereas 
a function as defined in Eqn.~\ref{eqn:ratio} 
results in a acceptable $\chi^2_\mathrm{red}(d.o.f.)=1.5(12)$ with
$\Delta\alpha=0.35\pm0.08$ and $c=(0.12\pm0.03)$~TeV$^{-1}$. 
We note that constraining the fit region to 
energies above $2$~TeV where absorption starts to become important, a pure power law
fit is not excluded any more for the ratio.

There is an overlap of the cut-off energies determined for Mkn~421 
($E_0=$\newcut)
and Mkn~501 ($E_0=$\oldcut) \citep{1999A&A...349...11A} taking the combined systematic and statistical errors into
account. 
However, the origin of the systematic errors  are very similar for both results
and therefore strongly correlated. As mentioned in Sect.~\ref{sect:systematics},
a 6\,\% systematic uncertainty on the relative energy scale remains  when comparing the two energy spectra.
Combining the independent statistical uncertainty and the estimated systematic uncertainty on
the relative energy scale results in
 $\Delta E=2.6\pm0.6_\mathrm{stat}\pm0.6_\mathrm{sys}$~TeV for the difference on the energy of the cut-off position.
  A detailed investigation 
of possible differences of energy spectra will be the issue of a forthcoming paper.

\section{Multi-wavelength observations}
\label{MWL}
  The flux increase observed from Mkn~421 in the TeV energy band has been accompanied by a strongly increased X-ray flux. 
Detailed studies of the correlation of the variability in both
energy bands are under-way. The simultaneously 
taken data from January/February 2001 with RXTE and HEGRA \citep{ICRC2001...horns} reveal a largely different spectral evolution for Mkn~421 in the X-ray and TeV being almost opposite to what has been observed from Mkn~501:
 Whereas the shape of the TeV-spectrum  of Mkn~501 remained constant during the 1997 activity period
\citep{1999A&A...349...11A}, the
Mkn~421 spectrum showed a spectral hardening with increasing flux.
A different spectral evolution is also seen 
in the X-ray spectrum for the two sources:
Whereas the X-ray spectrum of Mkn~501 is very hard during the flare which
is commonly interpreted as a 
shift of the synchrotron
peak position $E^{Sy}_{peak}$ from below 1~keV to 100~keV and back \citep{2001ApJ...554..725T}, the
X-ray spectrum of Mkn~421 remains soft with a 
 moderate shift of the synchrotron peak
from below 1~keV at a flux level of $\approx10^{-10}$~erg/(cm$^2$\,s)   \citep{2000MNRAS.312..123M} to $\approx 2-3$~keV
on January 31,2001 when the  flux measured between 2-10~keV 
reached $F_{2-10}=1.6\cdot10^{-9}$~erg/(cm$^2$\,s) \citep{ICRC2001...horns}.
These observations confirm the energy dependence of the peak-position claimed
to follow a power law behaviour with $E^{sy}_{peak}\propto F_{Sy}^{0.5}$ for
Mkn~421.  For Mkn~501 the relation of the peak position and flux follows 
$E^{sy}_{peak}\propto F_{Sy}^{2}$  underlining the 
very different  behaviour of the two extreme BL~Lac objects
\citep{2000ApJ...541..166F}.  

The TeV data  of Mkn~421 taken simultaneously with the X-ray observations in January/February 2001
 indicate a hard spectrum with a photon index 
$\alpha\approx2$ that deviate substantially from the 
quiescent energy spectrum being much softer ($\alpha\approx3$).
 The constraints on a one-zone SSC model of the source derived from
these multi-wavelength observations are severe. 
 The X-ray spectrum with a peak-energy $<3$~keV 
constrains the electron-spectrum which is responsible for the synchrotron peak emission to have a
low energy
($\gamma_{peak}=1.5\cdot10^{5}\delta_{50}^{-0.5}\,B_{0.2}^{-0.5}$).  
As a result,  without invoking large $\delta$ and a small magnetic field,
the SSC model predicts  a peak energy 
for the Compton-scattered spectrum below 1~TeV and the spectral shape is expected to be soft in the energy region from 1-20~TeV
because of the Klein-Nishina suppression of the Compton-scattering cross-section. 
However, the
\textit{observed} TeV-spectra are very hard and a small magnetic field $B<0.22$~G 
and a large Doppler-factor $\delta>50$ are required to achieve an adequate fit of the TeV data with a SSC-model (see e.g. \citet{2001ApJ...559..187K}). The absorption of TeV photons 
due to pair production with photons of the extragalactic radiation field
adds to the difficulties, because the source spectra 
corrected for the absorption are intrinsically harder than the observed
spectra which are not easily described with
a simple SSC-one-zone model (see e.g. \citet{felix_icrc}).

\section{Summary and conclusions}
   The observations of Mkn~421 at TeV-energies indicate significant spectral variations for different
flux levels. The photon index $\alpha$ changes from $\approx3$ to $\approx2$ with a
flux increasing from $F(E>1~\mathrm{TeV})=0.5\cdot F_{-11}\ldots10\cdot F_{-11}$. 
Diurnal flux variations have been observed with rise and decay times as short as $20$~min.
During particular nights where a complete flare was observed, 
the photon index determined during the preflare and the flare period differ by
as much as $\Delta \alpha=0.75\pm0.25$ getting harder while the flux increases.
 All energy spectra are well-fit by a power law with an exponential cut-off at $E_0=$\newcut\, without
indications for variations of the cut-off energy at different flux-levels. However, the sensitivity
on changes of the cut-off energy is limited by strong correlations between the two parameters ($\alpha$,$E_0$) 
used to describe the energy spectra. A comparison of $E_0$ for Mkn~501 and Mkn~421
indicates that the position of the cut-off energy for Mkn~501 is larger by \dcut\,
taking the combined systematic and statistical error into account. A difference of the cut-off energies for
two sources at similar red shift precludes the possibility of the cut-off being only an absorption feature.

\begin{acknowledgements}
The support of the German ministry for research and
technology (BMBF) and of the Spanish Research Council (CICYT) is gratefully
acknowledged. GR acknowledges receipt of a Humboldt fellowship.
We thank the Instituto de Astrof\'{\i}sica de Canarias
for the use of the site and for supplying excellent working conditions at
La Palma. 
\end{acknowledgements}

\end{document}